\documentclass[twocolumn,floatfix,superscriptaddress,preprintnumbers,nofootinbib,10pt]{revtex4}

\usepackage{graphicx}
\usepackage{color}

\newcommand{\bea}{\begin{eqnarray}}
\newcommand{\beq}{\begin{equation}}
\newcommand{\eea}{\end{eqnarray}}
\newcommand{\eeq}{\end{equation}}

\begin{document}
\preprint{TUM-HEP-746/10}

\title{\boldmath A Lower Bound on hadronic EDMs from CP Violation in $D^0-\bar D^0$
mixing in SUSY Alignment Models}

\author{Wolfgang~Altmannshofer}
\affiliation{Physik-Department, Technische Universit\"at M\"unchen,
James-Franck-Stra\ss e, D-85748 Garching, Germany}

\author{Andrzej~J.~Buras}
\affiliation{Physik-Department, Technische Universit\"at M\"unchen,
James-Franck-Stra\ss e, D-85748 Garching, Germany}
\affiliation{TUM Institute for Advanced Study, Technische Universit\"at M\"unchen,
\\Arcisstr.~21, D-80333 M\"unchen, Germany}

\author{Paride~Paradisi}
\affiliation{Physik-Department, Technische Universit\"at M\"unchen,
James-Franck-Stra\ss e, D-85748 Garching, Germany}

\begin{abstract}

The SM predictions for CP violating effects in $D$ meson systems are highly
suppressed at the per mill level. Therefore, any experimental evidence for
a sizable CP violation in $D^0-\bar D^0$ mixing would unambiguously point
towards a New Physics (NP) signal. Within supersymmetric scenarios, the
popular alignment models can naturally account for large, non-standard
effects in $D^0-\bar D^0$ mixing. We demonstrate that, within alignment
models, detectable CP violating effects in $D^0-\bar D^0$ mixing would
unambiguously imply a lower bound for the electric dipole moment (EDM)
of hadronic systems, like the neutron EDM and the mercury EDM, in the
reach of future experimental sensitivities. The simultaneous evidence of
CP violation in $D$ meson systems together with non vanishing hadronic EDMs
would strongly support the idea of SUSY alignment models and disfavour
gauge-mediated SUSY breaking models, SUSY models with MFV and non-Abelian
SUSY flavour models. As a byproduct of our analysis, we study the
correlation between the time dependent CP asymmetry in decays to CP 
eigenstates $S_f$ and the semileptonic asymmetry $a_{\rm SL}$ both model 
independently and in SUSY alignment models.

\end{abstract}

\maketitle

\section{Introduction}

The meson systems are among the most powerful low energy probes of New Physics (NP) and
can be regarded as {\it golden channels} of the {\it high intensity frontier}. However,
the $K$ and $B_d$ systems have been already well studied experimentally and all the
currently available measurements are compatible with the Standard Model (SM) predictions,
although large room for NP, in particular in rare $K$ decays, is still allowed. Similarly
the question of the size of NP in the $B_s$ system, in particular in CP-violating
observables and in rare decays like $B_s\to\mu^+\mu^-$ is still to be answered.

Yet, already the present data imply the so-called NP flavour and CP problems, that amount
to the tension between the solution of the hierarchy problem, requiring a TeV scale NP,
and the explanation of the Flavour Physics data in which this NP did not show up convincingly.

Within supersymmetric scenarios, the most popular mechanisms accounting for a solution of
the flavour problem are~\footnote{For a recent comprehensive analysis of flavour and CP
violating effects in SUSY theories see~\cite{Altmannshofer:2009ne}.}
\begin{itemize}
\item {\em MFV}. Flavour violation is assumed to be entirely described by the
CKM matrix even in theories beyond the SM~\cite{D'Ambrosio:2002ex,Buras:2000dm}.
\item {\em Decoupling}. The sfermion mass scale is taken to be very high.
\item {\em Degeneracy}. The sfermion masses are degenerate to a large extent
leading to a strong GIM suppression, as in (pure) gauge-mediated SUSY breaking
models~\cite{Giudice:1998bp} or non-Abelian SUSY flavour models~\cite{nonabelian}.
\item {\em Alignment}. The down-quark and squark mass matrices are aligned,
so that the flavour-changing down fermions-gaugino-sfermion couplings are
suppressed~\cite{Nir:1993mx,Leurer:1993gy}
\end{itemize}
%

In contrast, the SUSY CP problem is not addressed by the MFV principle or flavour
symmetries as the {\it flavour blind} phases generating unacceptably large EDMs
are not forbidden by the flavour
symmetry~\cite{Nir:1996am,Ross:2004qn,Colangelo:2008qp,Paradisi:2009ey,Kagan:2009bn}.
For instance, the extra assumption employed by SUSY flavour models is that CP is preserved 
by the sector responsible for SUSY breaking, while it is spontaneously broken in
the flavour sector~\cite{Nir:1996am,Ross:2004qn}.
In this case, one may expect that {\it flavoured} EDMs are generated~\cite{Hisano:2008hn}
and highly correlated with CP violating phenomena in flavour physics.


In this work, we focus on alignment models that naturally arise in the context
of Abelian flavour symmetries~\cite{Nir:1993mx,Leurer:1993gy}.
Such symmetries, can simultaneously explain the pattern of fermion masses and mixings and 
provide the sufficient suppression of FCNC phenomena~\cite{Nir:1993mx}. On the other hand,
a prominent feature of this class of models is the appearance of large FCNC effects in the
up-quark sector. Therefore, physical observables generated by $D^0-\bar D^0$ mixing provide
a unique tool to probe or to falsify models with alignment~\cite{Nir:1993mx,Leurer:1993gy,Nir:2007ac,Golowich:2007ka,Gedalia:2009kh}.

On general grounds, the $D$ system (as well as the $B_s$ system) offers a splendid
opportunity to discover CP violating effects arising from NP~\cite{Blaylock:1995ay,Bianco:2003vb,Nir:2005js,Grossman:2006jg} 
as the SM predictions are completely negligible, at the level of ${\cal O}((V_{cb}V_{ub})/(V_{cs}V_{us}))\sim10^{-3}$. As a consequence, any experimental
signal of CP violation in $D^0-\bar D^0$ above the per mill level would unambiguously 
point towards a NP effect.

In this context, the main goals of this letter are:
\begin{itemize}
\item[{\bf i)}] to quantify the NP room left to CP violation in $D^0-\bar D^0$ transitions
that is compatible with all the available experimental data
(see also~\cite{Grossman:2009mn,Bigi:2009df,Kagan:2009gb});
\item[{\bf ii)}] to point out strategies enabling to probe or to falsify NP scenarios
by means of a correlated analysis of low energy observables.
In particular, we point out that, within alignment models, CP violating effects in
$D^0-\bar D^0$ mixing unambiguously imply a lower bound for the EDMs of hadronic
systems, like the neutron EDM and the mercury EDM, in the reach of the future experimental sensitivities.
\end{itemize}

As a byproduct of our study of CP violation in $D^0-\bar D^0$ mixing, we also analyse the
correlation between the time dependent CP asymmetry in decays to CP eigenstates $S_f$ and
the semileptonic asymmetry $a_{\rm SL}$ both model independently and in SUSY alignment models.

\section{\boldmath Model independent considerations on $D^0-\bar D^0$ mixing}

We start with a brief summary of the basic formalism of $D^0-\bar D^0$ mixing focusing on
those aspects most relevant for our analysis. Recent comprehensive presentations can be
found e.g. in \cite{Grossman:2009mn,Bigi:2009df,Kagan:2009gb}. In the following we use
the conventions of~\cite{Bigi:2009df}.

The amplitude for the $D^0-\bar D^0$ transition can be written as 
\begin{eqnarray}
\langle D^0 |\mathcal{H}_{\rm eff}| \bar D^0 \rangle &=& M_{12} - \frac{i}{2} \Gamma_{12} ~, \\
\langle \bar D^0 |\mathcal{H}_{\rm eff}| D^0 \rangle &=& M_{12}^* - \frac{i}{2} \Gamma_{12}^* ~,
\end{eqnarray}
where $M_{12}$ is the dispersive part and $\Gamma_{12}$ the absorptive part. The phases
of $M_{12}$ and $\Gamma_{12}$ are phase convention dependent but their relative phase is
a physical observable. The three fundamental theory parameters describing $D^0-\bar D^0$
mixing are then
\begin{equation}
|M_{12}| ~,~~ |\Gamma_{12}| ~,~~ \phi_{12} = \textnormal{Arg}(M_{12}/\Gamma_{12})~.
\end{equation}

The neutral $D$ meson mass eigenstates $D_1$ and $D_2$ are linear combinations of
the strong interaction eigenstates, $D^0$ and $\bar D^0$
\begin{equation}
| D_{1,2} \rangle = p | D^0 \rangle \pm q | \bar D^0 \rangle ~,
\end{equation}
with the convention CP $|D^0\rangle = + |\bar D^0\rangle$ and
\begin{equation}
\label{q/p}
\frac{q}{p} = \sqrt{\frac{M_{12}^* - \frac{i}{2} \Gamma_{12}^*}{M_{12} - \frac{i}{2} \Gamma_{12}}} ~.
\end{equation}
Their normalized mass and width differences, $x$ and $y$, are given by
\begin{equation}
x = \frac{\Delta M_D}{\Gamma} = 2 \tau \textnormal{Re}\left[ \frac{q}{p} \left( M_{12} - \frac{i}{2} \Gamma_{12} \right) \right] ~,
\end{equation}
\begin{equation}
y = \frac{\Delta \Gamma}{2 \Gamma} = - 2 \tau \textnormal{Im}\left[ \frac{q}{p} \left( M_{12} - \frac{i}{2} \Gamma_{12} \right) \right] ~,
\end{equation}
with the lifetime of the neutral $D$ mesons $\tau = 1/\Gamma = 0.41$ps.

Experimentally, $D^0 - \bar D^0$ mixing is now firmly established with the non-mixing
hypothesis $x=y=0$ excluded at $10.2\sigma$~\cite{Barberio:2008fa,Schwartz:2009jv}.
Still, at the current level of sensitivity, there is no evidence for CP violation in
$D^0-\bar D^0$ mixing. The experimental data on both $|q/p|$ and $\phi=\textnormal{Arg}(q/p)$
is compatible with CP conservation, i.e. $|q/p|=1$ and $\phi=0$. The most recent world
averages as obtained by HFAG read~\cite{Barberio:2008fa,Schwartz:2009jv}
\begin{eqnarray} 
\label{eq:x_bound}
x &=& (0.98^{+0.24}_{-0.26})\% ~, \\ 
\label{eq:y_bound}
y &=& (0.83 \pm 0.16)\% ~, \\ 
\label{eq:qp_bound}
\left| q/p \right| &=& 0.87^{+0.17}_{-0.15} ~, \\ 
\label{eq:phi_bound}
\phi &=& (-8.5^{+7.4}_{-7.0})^\circ ~.
\end{eqnarray}
Throughout our analysis, we use these constraints at the $2\sigma$ level.

In the remainder of our analysis of CP violation in $D^0 - \bar D^0$ mixing we
will focus on two observables: the time dependent CP asymmetry in decays to CP
eigenstates $S_f$ and the semileptonic asymmetry $a_{\rm SL}$.

\subsection{Time dependent CP asymmetry}

\begin{figure*}[th]
\includegraphics[width=0.31\textwidth]{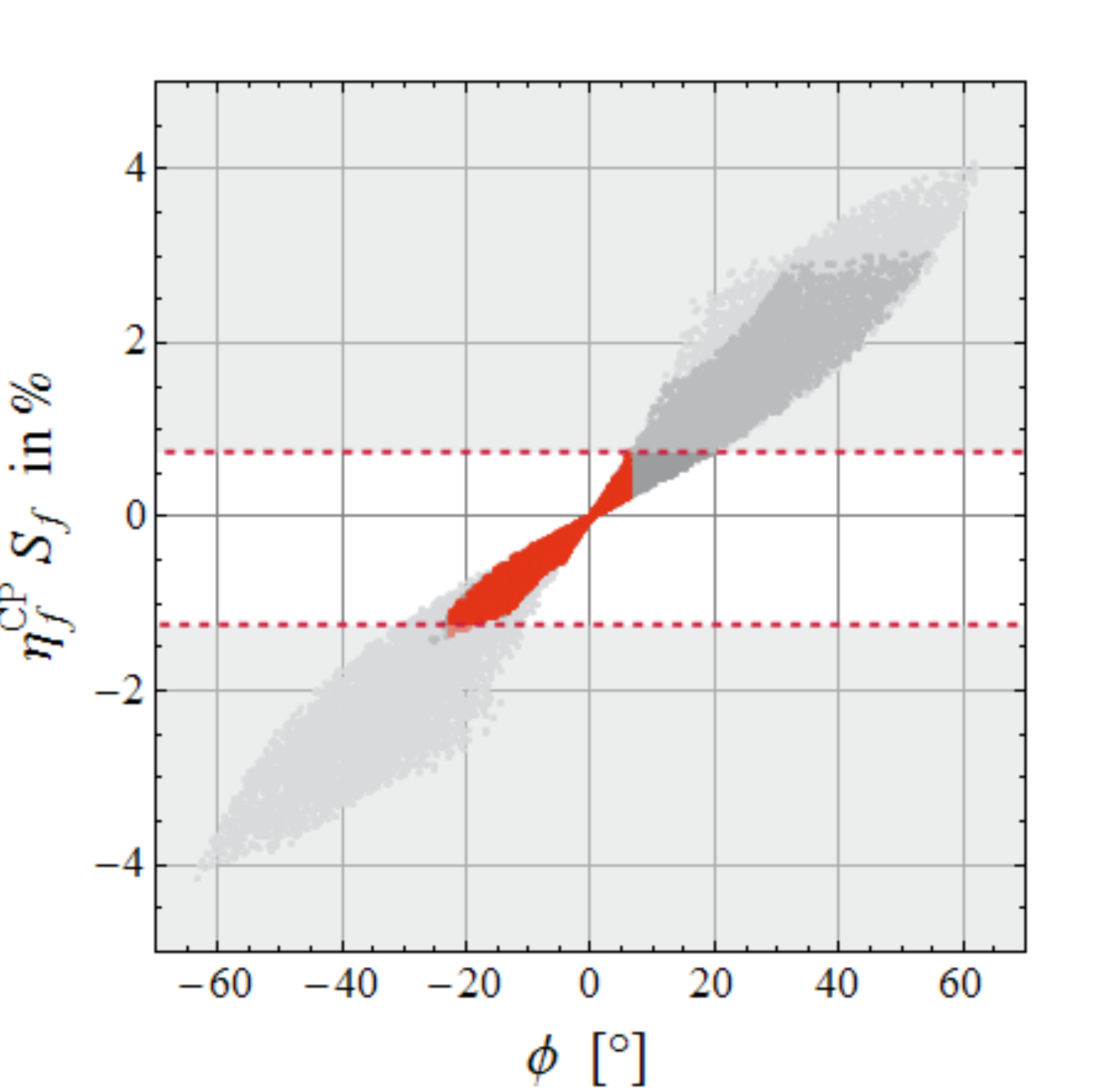}~~~
\includegraphics[width=0.31\textwidth]{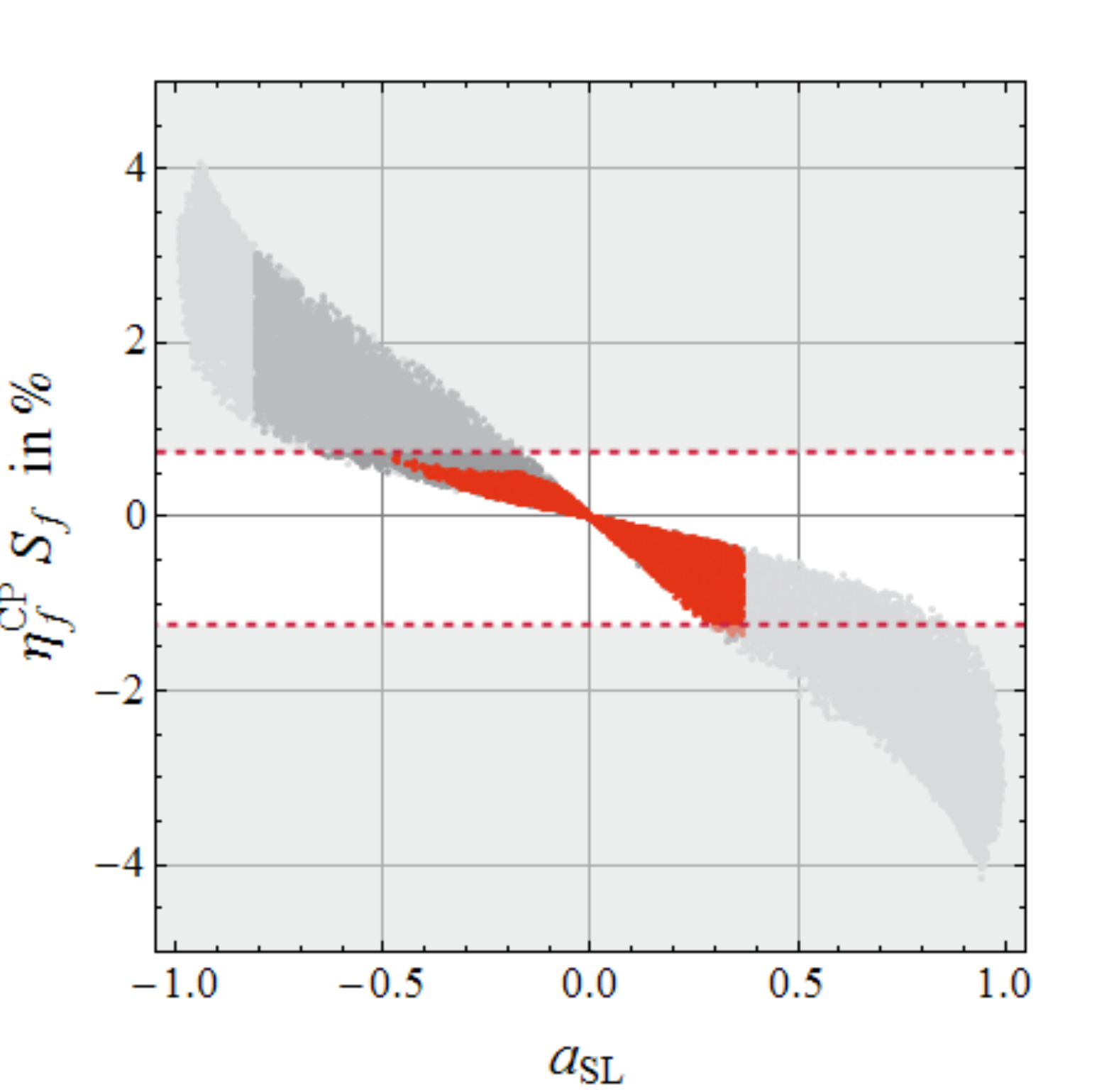}
\caption{Model independent correlations between $S_f$ and $\phi$ (left) and
$a_{\rm SL}$ and $S_f$ (right).
Light gray points satisfy the constraints (\ref{eq:x_bound}) and (\ref{eq:y_bound}) from $x$
and $y$ while darker gray points further satisfy the constraint (\ref{eq:qp_bound}) from $|q/p|$.
Red points are compatible with all constraints listed in (\ref{eq:x_bound}) - (\ref{eq:phi_bound})
and the dashed lines stand for the allowed range~(\ref{eq:DYf_bound}) for $\eta^{\rm CP}_{f}S_f$.}
\label{fig:model_independent}
\end{figure*}

The decay rates of neutral $D$ mesons decaying to CP eigenstates $f$ are to a good
approximation given by~\cite{Bergmann:2000id,Nir:2005js,Kagan:2009gb}
\begin{eqnarray}
\Gamma(D^0(t) \to f) &\propto& \text{exp} \left[-\hat\Gamma_{D^0 \to f}t\right]~, \\
\Gamma(\bar D^0(t)\to f) &\propto&\text{exp}\left[-\hat\Gamma_{\bar D^0\to f}t\right]~,
\end{eqnarray}
with effective decay widths
\begin{eqnarray}
\hat \Gamma_{D^0 \to f} &=&
\Gamma_D\left[1+\eta_f^{\rm CP}\left|\frac{q}{p}\right|\left(y\cos\phi-x\sin\phi\right)\right]~, \\
\hat\Gamma_{\bar D^0 \to f} &=& \Gamma_D \left[1 + \eta_f^{\rm CP} \left| \frac{p}{q} \right| \left( y\cos\phi + x\sin\phi \right) \right]~.
\end{eqnarray}
Here $\eta_f^{\rm CP}$ is the CP parity of the final state $f$.

One then defines the following CP violating
combination~\cite{Bergmann:2000id,Bigi:2009df,Kagan:2009gb}
\begin{equation}
S_f = 2 \Delta Y_f =
\frac{1}{\Gamma_D}\left(\hat\Gamma_{\bar D^0 \to f} - \hat\Gamma_{D^0 \to f}\right) ~,
\end{equation}
that is given by
\begin{equation} \label{eq:DYf}
\eta_f^{\rm CP} S_f = x \left(\left|\frac{q}{p}\right| + \left|\frac{p}{q}\right|\right)\sin\phi - y \left( \left|\frac{q}{p}\right| - \left|\frac{p}{q}\right| \right) \cos\phi ~.
\end{equation}
This expression gets in principle modified in the presence of new weak phases in the decay~\cite{Bigi:2009df,Kagan:2009gb}, that one could expect e.g. in singly Cabibbo
suppressed decay modes~\cite{Grossman:2006jg}.
However, as shown in \cite{Kagan:2009gb}, such effects are severely constrained by
the existing data on time integrated CP asymmetries. To an excellent approximation, 
eq.~(\ref{eq:DYf}) holds therefore also in the presence of new weak phases in the decay,
i.e. $\eta_f^{\rm CP} S_f $ and $\eta_f^{\rm CP}\Delta Y_f$ are universal for all final
states and practically independent of direct CP violation in the decays.

In fact time dependent CP asymmetries are currently determined from the singly Cabbibo suppressed
decay modes $D^0 \to K^+ K^-$ and $D^0 \to \pi^+ \pi^-$ and one has~\cite{Barberio:2008fa}
\begin{equation}\label{eq:DYf_bound}
\eta_f^{\rm CP} S_f = (-0.248 \pm 0.496)\%~.
\end{equation}
Concerning Cabbibo favored decay modes, the most promising channel seems to be
$D^0 \to K_S \phi$~\cite{Bigi:2009df}. The corresponding time dependent CP asymmetry
has been studied within the Littlest Higgs model with T-parity in~\cite{Bigi:2009df}
and within a Randall-Sundrum framework in \cite{Bauer:2009cf}.

In the left plot of fig.~\ref{fig:model_independent} we show the dependence of the asymmetry 
$\eta^{\rm CP}_{f}S_f$ on $\phi$. The light gray points only fulfill the constraints from $x$
and $y$, darker gray points in addition also the constraint from $|q/p|$. Red points finally
are compatible with all constraints listed in (\ref{eq:x_bound}) - (\ref{eq:phi_bound}).
The range~(\ref{eq:DYf_bound}) for $\eta^{\rm CP}_{f}S_f$ is also shown as dashed lines.
We observe a very significant impact of the constraint on $\phi$ in (\ref{eq:phi_bound}),
that has not been taken into account in \cite{Bigi:2009df} in order to be conservative.
In fact, our analysis emphasizes the crucial role played by a precise measurement of $\phi$ 
to establish the NP room left to CP violation in D-physics.

To obtain this plot we proceed in the following way: while the absorptive part of the
$D^0 - \bar D^0$ mixing amplitude is not sensitive to new short distance dynamics, the
dispersive part can be affected by NP effects. We therefore decompose the mixing amplitude
as follows
\begin{eqnarray}
M_{12} &=& M_{12}^{\rm SM} + M_{12}^{\rm NP}~, \\
\Gamma_{12} &=& \Gamma_{12}^{\rm SM}~.
\end{eqnarray}
Both SM contributions, $M_{12}^{\rm SM}$ and $\Gamma_{12}^{\rm SM}$ are predicted to be real
(in the CKM convention) to an excellent approximation, but their magnitude cannot be calculated
in a reliable way~\cite{Falk:2001hx,Falk:2004wg}. We follow~\cite{Ciuchini:2007cw,Bauer:2009cf}
and scan $M_{12}^{\rm SM}$ flatly in the range $[-0.02 , 0.02] {\rm ps}^{-1}$, so
that the SM contribution alone can saturate the experimental bound. Moreover,
we scan $\Gamma_{12}^{\rm SM}$ flatly in the range $[-0.04 , 0.04] {\rm ps}^{-1}$.

For the model independent analysis we then scan the NP contribution with
$0 < |M_{12}^{\rm NP}| < 0.05 {\rm ps}^{-1}$ and $0 < {\rm Arg}(M_{12}^{\rm NP}) < 2\pi$.

\subsection{Semileptonic Asymmetry}

The semileptonic asymmetry in the decay to ``wrong sign'' leptons is defined as
\begin{eqnarray}
a_{\rm SL} &=& \frac{\Gamma(D^0 \to K^+ \ell^- \nu) - \Gamma(\bar D^0 \to K^- \ell^+ \nu)}{\Gamma(D^0 \to K^+ \ell^- \nu) + \Gamma(\bar D^0 \to K^- \ell^+ \nu)} \nonumber \\
&=& \frac{|q|^4-|p|^4}{|q|^4+|p|^4}
\end{eqnarray}
and is a direct measure of CP violation in the mixing.

The model independent relation between $x$, $y$, $|q/p|$ and $\phi$ identified in~\cite{Grossman:2009mn} implies the following correlation between the universal
time dependent CP asymmetry $S_f$ and the semileptonic asymmetry
$a_{\rm SL}$~\cite{Bigi:2009df,Kagan:2009gb}
\begin{equation}
\label{eq:DYf_vs_aSL}
S_f = 2 \Delta Y_f = - \eta_f^{\rm CP}~\frac{x^2+y^2}{|y|}~a_{\rm SL}~.
\end{equation}
In the right plot of fig.~\ref{fig:model_independent} we show exactly this correlation as
a result of the numerical scan described above. We observe that the constraints coming
from $|q/p|$ and $\phi$ limit the allowed range for $a_{\rm SL}$ in a model independent way.
The remaining range reads $-0.5 \lesssim a_{\rm SL} \lesssim 0.35$~.

\section{\boldmath $D^0-\bar D^0$ mixing in SUSY alignment models}

The quark-squark alignment mechanism
occurs naturally in models with Abelian horizontal symmetries reproducing the
observed hierarchy in the Yukawa couplings.
They make use of holomorphic zeros in the down quark mass matrix to obtain
the desired suppression for the mixing angles of the first two generations.
As shown in Ref.~\cite{Nir:2002ah}, in the framework of alignment it is possible 
to predict for a broad class of Abelian flavour models 
both lower and upper bounds for the SUSY flavour mixing angles $(K^q_M)_{ij}$ 
that parameterize the $\tilde g-q_{M_i}-\tilde q_{M_j}$ couplings, with $M = 
L, R$ (see tab.~\ref{QSA}).

\begin{table}[tb]
\begin{tabular}{c|c|c}
Mixing Angle  &  Lower Bound  & Upper Bound  \\
\hline\hline
$(K^d_{L})_{12}$ & $\lambda^5$ & $\lambda^3$ \\
$(K^d_{R})_{12}$ & $\lambda^7$ & $\lambda^3$ \\
\hline
$(K^d_{L})_{13}$ & $\lambda^3$ & $\lambda^3$ \\
$(K^d_{R})_{13}$ & $\lambda^7$ & $\lambda^3$ \\
\hline
$(K^d_{L})_{23}$ & $\lambda^2$ & $\lambda^2$ \\
$(K^d_{R})_{23}$ & $\lambda^4$ & $\lambda^2$ \\
\hline
$(K^u_{L})_{12}$ & $\lambda$   & $\lambda$ \\
$(K^u_{R})_{12}$ & $\lambda^4$ & $\lambda^2$ \\
\hline\hline
\end{tabular}\vspace*{4pt}
\caption{Lower and upper bounds on SUSY flavour mixing angles in alignment models
as reported in Ref.~\cite{Nir:2002ah}.}
\label{QSA}
\end{table}

The most peculiar feature of alignment models is the appearance of a large 
mixing angle $(K^u_{L})_{12}\sim\lambda$, leading to large effects in $D^0 - 
\bar D^0$ mixing. This can be understood in the following way. 
Abelian flavour symmetries do not impose any restriction on the mass
splittings between squarks of different generations therefore they are expected to be
non-degenerate with natural order one mass splittings. In particular, in the presence
of a mass splitting between the first two generations of left-handed squarks, the
(basis-independent) $SU(2)_L$ relation between the left-left blocks of 
the up- and down-squark mass matrices
$\tilde M^{2u}_{LL}= V \tilde M^{2d}_{LL} V^{\dagger}$ implies that
\begin{equation}
\label{qsa_approx}
(\tilde M^{2u}_{LL})_{21} =
\left[V \tilde M^{2d}_{LL}V^{\dagger}\right]_{21}
\simeq
(\tilde M^{2d}_{LL})_{21} + \lambda\left(\tilde m_2^2-\tilde m_1^2\right)~.
\end{equation}
The last relation holds in the CKM convention at the first order in the expansion parameter
$\lambda$. Thus, even for $(\tilde M^{2d}_{LL})_{21} = 0$, which is approximately satisfied
in alignment models, there are irreducible flavour violating terms in the up squark sector
driven by the CKM as long as the left-handed squarks are split in mass.

The corresponding SUSY contributions to $D^0-\bar D^0$ mixing can be obtained, as usual, by
means of the following $\Delta F = 2$ effective Hamiltonian
\begin{equation}
\mathcal{H}_{\rm eff} = \sum_{i=1}^5 C_i Q_i + \sum_{i=1}^3 \tilde C_i \tilde Q_i ~+{\rm h.c.}~,
\end{equation}
with the operators $Q_i$ given by
\begin{eqnarray}
Q_1 & = &
(\bar u^\alpha \gamma_\mu P_L c^\alpha)(\bar u^\beta \gamma^\mu P_L c^\beta)~,\nonumber \\
Q_2 & = &
(\bar u^\alpha P_L c^\alpha)(\bar u^\beta P_L c^\beta)~,\nonumber \\
Q_3 & = &
(\bar u^\alpha P_L c^\beta)(\bar u^\beta P_L c^\alpha)~,\nonumber \\
Q_4 & = &
(\bar u^\alpha P_L c^\alpha)(\bar u^\beta P_R c^\beta)~,\nonumber \\
Q_5 & = &
(\bar u^\alpha P_L c^\beta)(\bar u^\beta P_R c^\alpha)~,
\end{eqnarray}
where $P_{R,L}=\frac{1}{2}(1\pm\gamma_5)$ and $\alpha,\beta$ are colour indices.
The operators $\tilde{Q}_{1,2,3}$ are obtained from $Q_{1,2,3}$ by the replacement
$L \leftrightarrow R$.
The dominant contributions to the Wilson coefficients come from gluino boxes, 
that, in the mass insertion approximation (MIA), are given at
LO in~\cite{Gabbiani:1996hi} and at NLO in~\cite{Ciuchini:2006dw}.

In the presence of large mass splittings among squarks, the MIA might not be 
a good method to calculate the SUSY contributions to FCNC processes. However, 
starting from eq.~(\ref{qsa_approx}),
the full computation can be reproduced quite accurately provided we define the MI
$(\delta_{u}^{L})_{21}$ as follows~\cite{Nir:2002ah}
\beq
\label{defdel}
(\delta_{u}^{L})_{21}\simeq\lambda~{\Delta\tilde m^2_{21}\over\tilde m^2_Q}~,
\eeq
where $\tilde m_Q=(\tilde m_2+\tilde m_1)/2$~\cite{Raz:2002zx}, $\Delta\tilde m^2_{21}=\tilde m_2^2-\tilde m_1^2$ and the CKM convention has been still assumed.

\begin{figure}[t]
\includegraphics[width=0.48\textwidth]{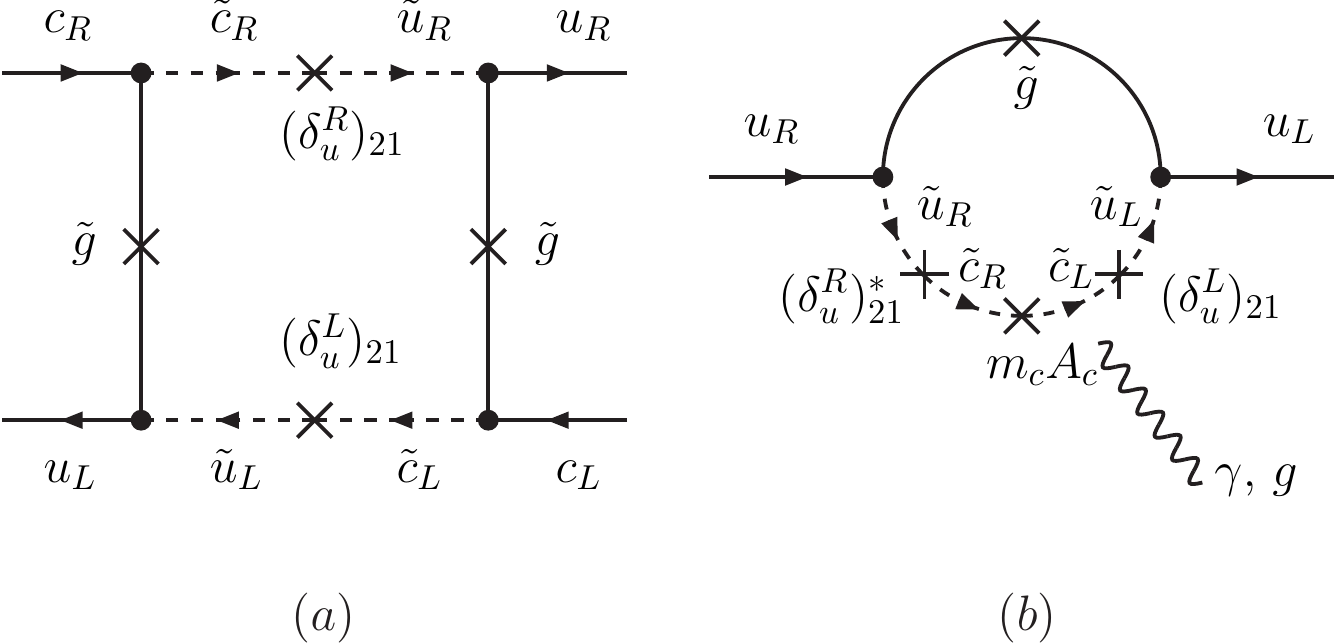}
\caption{Examples of relevant Feynman diagrams contributing $(a)$ to
$D^0 - \bar D^0$ mixing and $(b)$ to the up quark (C)EDM in SUSY alignment models.}
\label{fig:diagrams}
\end{figure}

We notice that, within alignment models, the CP violating WCs in the super-CKM basis
in the MIA are $\tilde{C}_{1}\sim (\delta_u^{R})_{21}(\delta_u^{R})_{21}$,
$(C_{4}, C_{5})\sim (\delta_u^{L})_{21}(\delta_u^{R})_{21}$,
while $C_{1}$ is real as $C_{1}\sim (\delta_u^{L})_{21}(\delta_u^{L})_{21}$
with $(\delta_u^{L})_{21}\sim\lambda$, given in~(\ref{defdel}).
Still, $C_{1}$ contributes dominantly to $\Delta M_D$. On the other hand, the dominant
contribution to ${\rm Im} M_{12}^{\rm NP}$, that is relevant for CP violating effects,
mostly arises from $C_{4}$ (see fig.~\ref{fig:diagrams}).
In particular, neglecting RGE-induced QCD effects, one can find the following rough
approximation for ${\rm Im}M_{12}^{\rm NP}$\footnote{In our numerical analysis instead
we use the full set of leading order SUSY contributions to the Wilson coefficients as
reported e.g. in~\cite{Altmannshofer:2007cs}, we include NLO RGE running according to \cite{Ciuchini:1997bw,Buras:2000if} and use the matrix elements from~\cite{Becirevic:2001xt,Ciuchini:2007cw}.}
\beq
\label{M^D_12}
{\rm Im}M_{12}^{\rm NP}
\simeq
{\alpha_s^2\over\tilde m_Q^2}
\frac{m_D^2}{m_c^2}m_Df_D^2
{7\over12}\tilde f_6(x_g)
{\rm Im}[(\delta_{u}^{L})_{21}(\delta_{u}^{R})_{21}],
\eeq
where $x_g = m_{\tilde g}^2/\tilde m_Q^2$ and $\tilde{f}_{6}(1)=1/20$.
The explicit expression for $\tilde{f}_{6}$ is given in~\cite{Gabbiani:1996hi}.
In (\ref{M^D_12}), we notice chiral enhancement $(m_D/m_c)^2$ that is stronger
than in B mixing but weaker than for the K system.

Finally, let us comment on possible constraints arising from FCNC processes involving
down quarks (like $K-\bar K$ mixing, $K\to\pi\nu\bar\nu$, ...). While one might expect
large NP effects in such processes from chargino-up squark contributions driven by the
large off-diagonal entries in $\tilde M^{2u}_{LL}$, we notice that chargino contributions
are actually sensitive to the off-diagonal elements of $\tilde M^{2d}_{LL}$ and not
$\tilde M^{2u}_{LL}$. This can be immediately seen remembering that the chargino induced
FCNC amplitude $A^{\tilde\chi}_{ij}$ is such that
$A^{\tilde\chi}_{ij}\sim (V^{\dagger}\tilde M^{2u}_{LL}V)_{ij}\equiv(\tilde M^{2d}_{LL})_{ij}$,
where we have used the $SU(2)_L$ relation $\tilde M^{2u}_{LL}= V \tilde M^{2d}_{LL} V^{\dagger}$.
Therefore, the huge MI $(\delta^{L}_{u})_{21}\sim \lambda$, arising in alignment models,
is not constrained by $K-\bar K$ observables.

\begin{figure*}[th]
\includegraphics[width=0.31\textwidth]{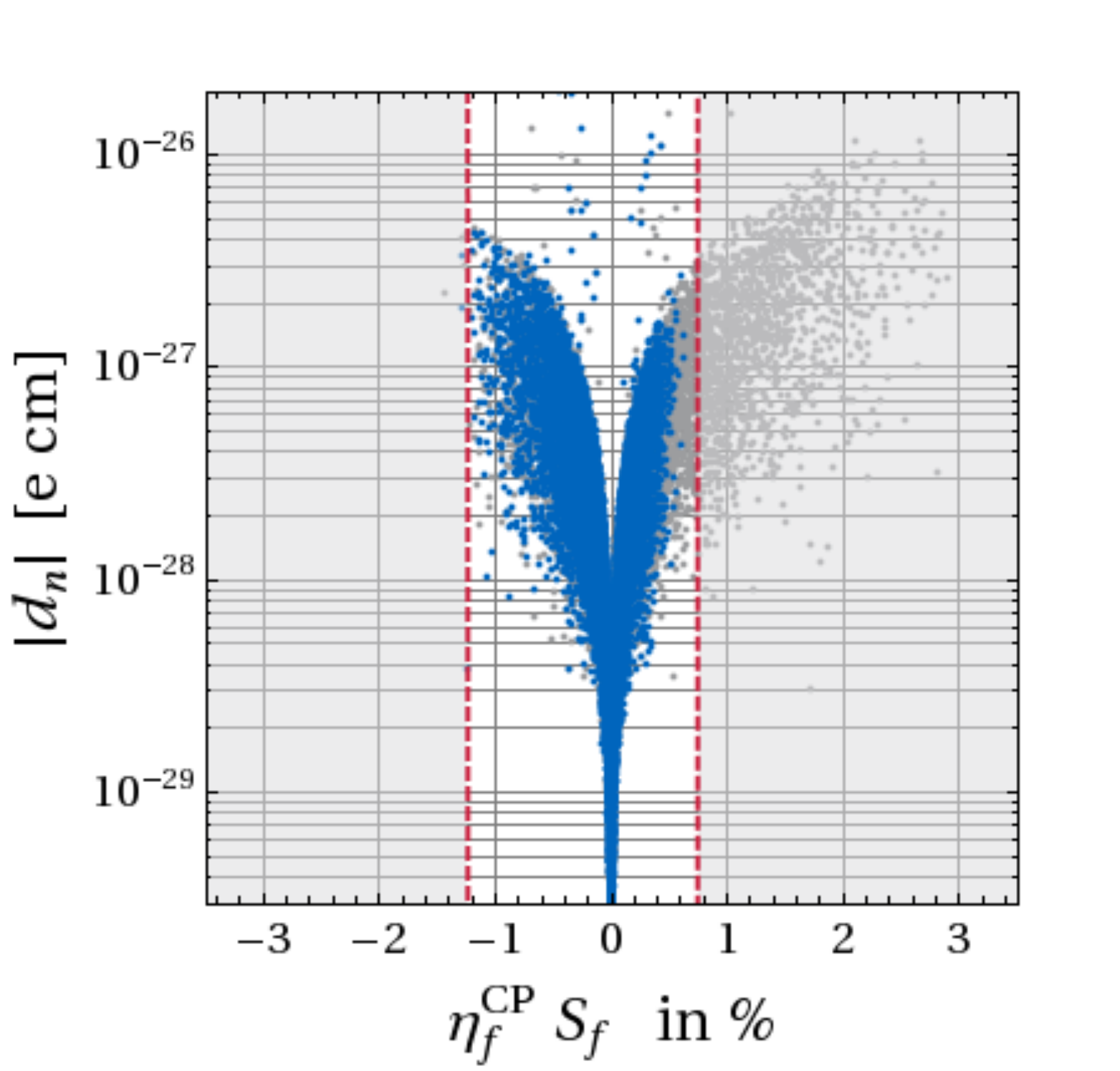}~~~
\includegraphics[width=0.31\textwidth]{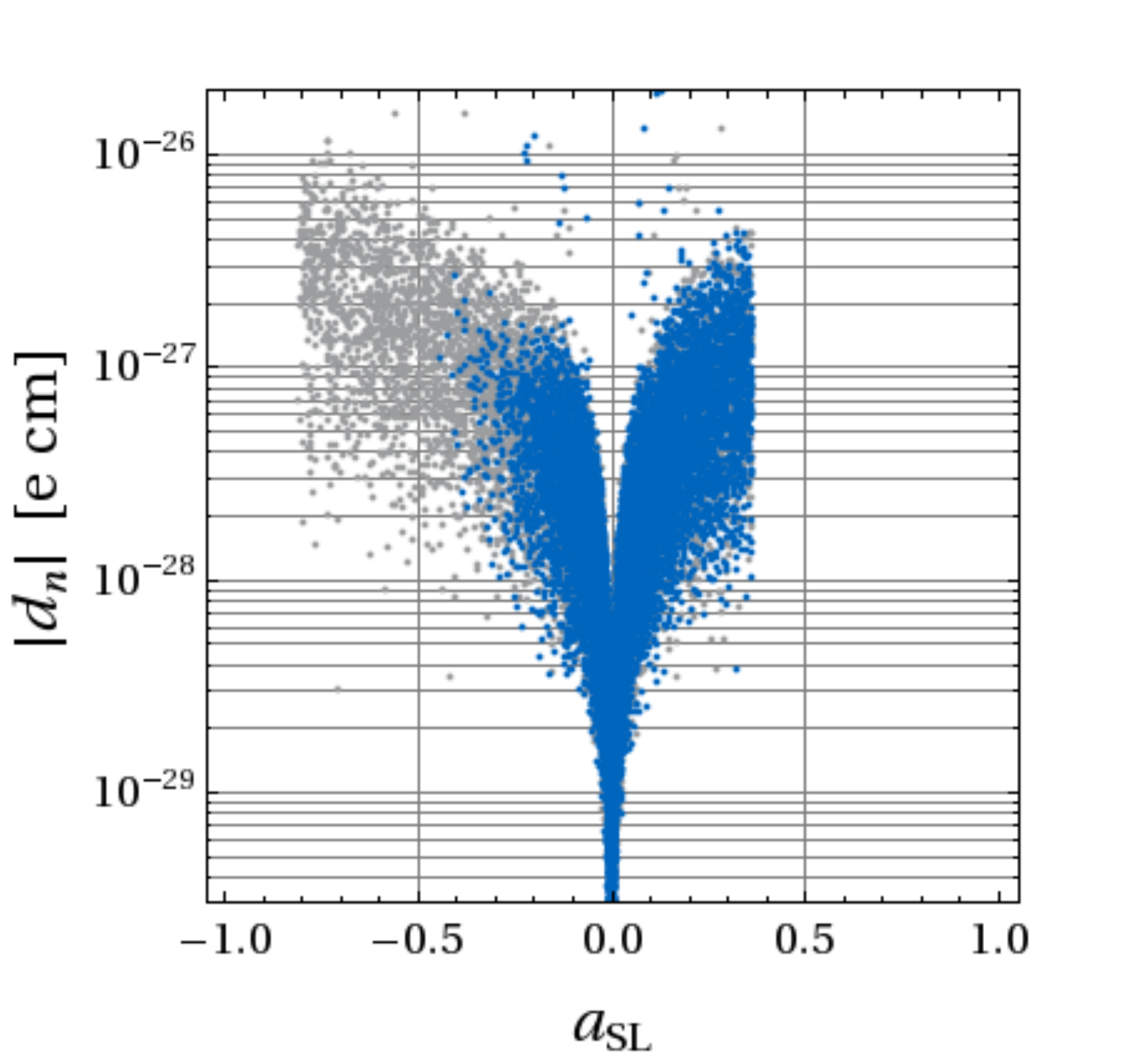}~~~
\includegraphics[width=0.30\textwidth]{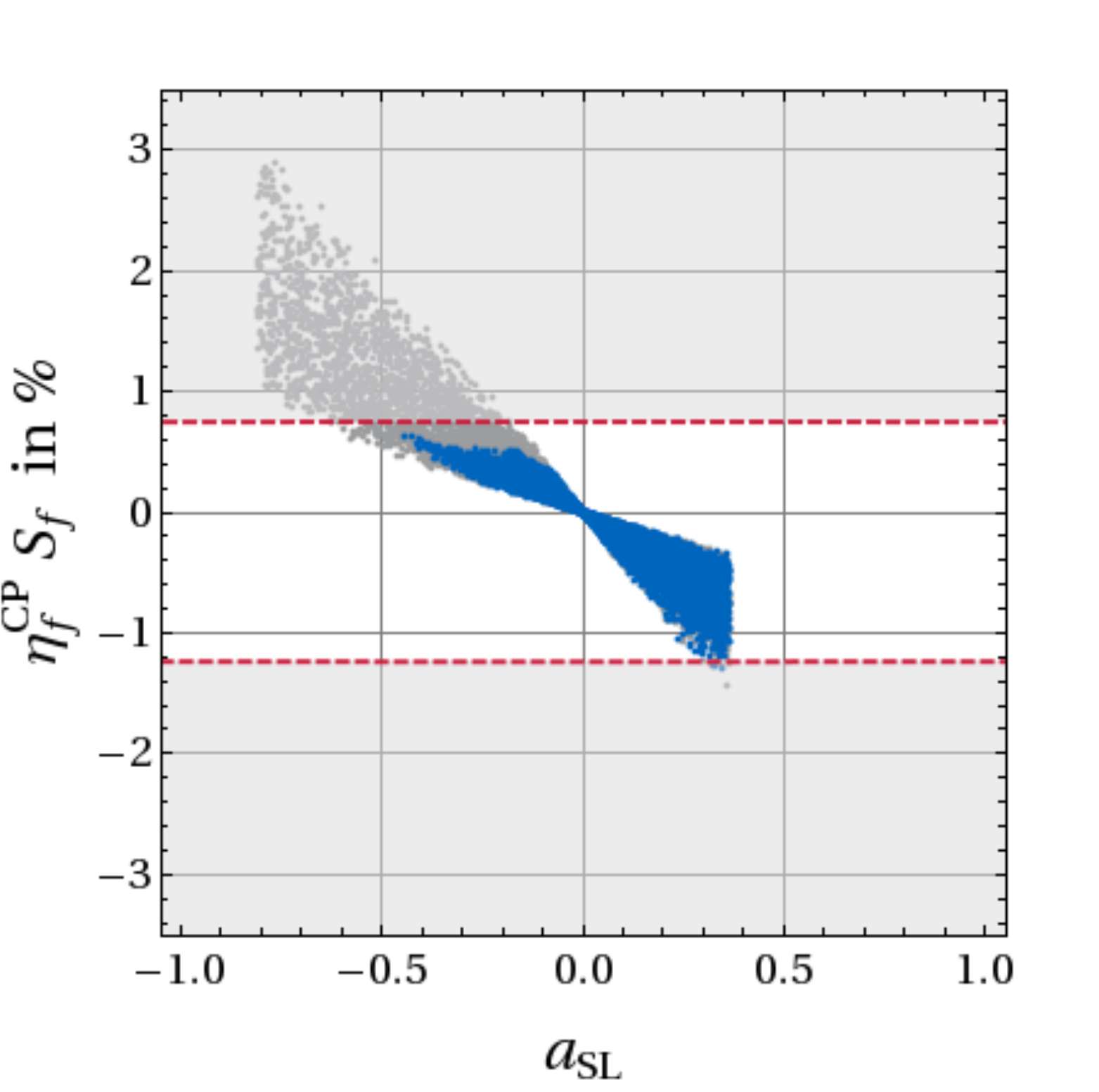}
\caption{Correlations between $d_n$ and $S_f$ (left), $d_n$ and $a_{\rm SL}$
(middle) and $a_{\rm SL}$ and $S_f$ (right) in SUSY alignment models.
Gray points satisfy the constraints (\ref{eq:x_bound})-(\ref{eq:qp_bound})
while blue points further satisfy the constraint (\ref{eq:phi_bound}) from $\phi$.
Dashed lines stand for the allowed range~(\ref{eq:DYf_bound}) for $S_f$.}
\label{fig:SUSY}
\end{figure*}

\section{Hadronic EDMs in SUSY alignment models}

The SM predictions for the electric dipole moments are very far from the present experimental
resolutions, therefore the EDMs represent very clean probes of NP effects. Within a MSSM
framework with flavour violating soft terms, large effects for the hadronic EDMs can be
naturally generated~\cite{Hisano:2008hn}. In particular, within alignment models, we find that
the dominant SUSY contributions to the hadronic EDMs arise from the gluino-squark contribution
to the up-quark (C)EDMs (see fig.~\ref{fig:diagrams}). Its expression at the SUSY scale reads
\beq
 \left\{ \frac{d_{u}}{e},~d^{c}_{u}\right\}\simeq
 -\frac{\alpha_s}{4\pi}\frac{m_{c}}{\tilde{m}^{2}_{Q}}
 \frac{m_{\tilde g} A_{c}}{\tilde{m}^{2}_{Q}} f(x_g)
 {\rm Im}\left[(\delta_{u}^{L})_{21}(\delta_{u}^{R})^{\star}_{21}\right]\,,
 \label{Eq:edm_u_gluino}
\eeq
where $A_{c}$ is the trilinear term relative to the second up-squark generation,
$f(1)= \{-8/135,~11/180\}$ and the explicit form for the loop function $f$ is
given in~\cite{Hisano:2008hn}.

We notice that, $d^{(c)}_{u}$ turns out to be proportional to $m_c$ instead of $m_u$,
as it happens in the case of {\it flavour blind} phases. This huge chiral enhancement
factor, stemming from flavour effects, can bring the hadronic (C)EDMs close to the
current and future experimental sensitivities, providing a splendid opportunity to
probe the flavour structure of the soft sector. Among the hadronic EDMs, the mercury
and the neutron EDMs, $d_{\rm Hg}$ and $d_n$ respectively, are especially important.
In particular, within alignment models, these EDMs can be expressed in terms of the
constituent quark EDMs as~\cite{Pospelov:2005pr,Raidal:2008jk}
\begin{eqnarray}
\label{Eq:dn_odd}
d_n &\simeq& (1\pm 0.5)\Big[-0.35\,d_u + e\,0.55\,d^c_u\Big]~, \\
\label{Eq:dHg}
d_{\rm Hg} &\simeq& 7 \times 10^{-3} \,e\,d^c_u~,
\end{eqnarray}
while the current experimental bounds on $d_n$~\cite{Baker:2006ts} and
$d_{\rm Hg}$~\cite{Griffith:2009zz} are, respectively
\begin{eqnarray}
\label{edm_dn_exp}
|d_n| &<& 2.9 \times 10^{-26}~e\,\rm{cm}~(90\% \rm{CL})\,,\\
\label{edm_dHg_exp}
|d_{\rm Hg}| &<& 3.1 \times 10^{-29}~e\,\rm{cm}~(95\% \rm{CL})\,.
\end{eqnarray}
Given that within SUSY alignment models $d_u \approx -d^c_u$, it turns out that
$|d_{\rm Hg}|\approx 10^{-2}~|d_n|$ hence, according to
eqs.~(\ref{Eq:dn_odd})-(\ref{edm_dHg_exp}), $|d_{\rm Hg}|$ would roughly be
one order of magnitude more sensitive than $d_n$ to SUSY effects. However, given
the large theoretical uncertainties affecting the prediction of $d_{\rm Hg}$ in
eq.~(\ref{Eq:dHg}), in the following, to be conservative, we focus only on $d_n$.

In the next section, we will show that a striking prediction arising within SUSY
alignment models is the prediction of a lower bound for $d_n$, in the reach of
the upcoming experimental resolutions, if non-standard CP violating effects in
$D^0-\bar D^0$ mixing are also generated.

\section{Numerical analysis}

As already discussed, SUSY alignment models can naturally satisfy the stringent bounds
from $\epsilon_K$ and $\Delta M_K$ by construction (see table~\ref{QSA}). In contrast,
large effects in $D^0-\bar D^0$ mixing are unambiguously predicted, provided the first
two squark families are non-degenerate, as we expect from naturalness arguments.

In our numerical analysis, we assume a CMSSM-like spectrum and perform a scan over the input
parameters: $m_0<2$~\rm{TeV}, $M_{1/2}<1$~\rm{TeV}, $|A_0|<3m_0$, and $5<\tan\beta<55$.
Moreover, we set at the GUT scale $(\delta^{R}_{u})_{21} = \lambda^3$ and the mass splitting
between the 1st and 2nd squark generation masses as $m_{\tilde{u}_L}=2 m_{\tilde{c}_L}=2m_0$.
Therefore, it turns out that $(\delta^{L}_{u})_{21}\sim \lambda$ at the GUT scale. However,
the above mass splitting is significantly reduced at the low scale because of a degeneracy
mechanism mostly driven by the flavour blind $SU(3)$
interactions~\cite{Dine:1990jd,Brignole:1993dj,Nir:2002ah}. As a result, the low scale $(\delta^{L}_{u})_{21}$ is significantly smaller than $\lambda$, typically by one order
of magnitude, and the constraints from $D$ mixing (\ref{eq:x_bound}) - (\ref{eq:phi_bound})
can be satisfied even for squark masses well below the TeV scale.

In the plot on the right of fig.~\ref{fig:SUSY}, we notice the correlation between
$S_f$ and $a_{\rm SL}$, as expected by the model-independent relation 
of eq.~(\ref{eq:DYf_vs_aSL}). Interestingly, within alignment models, it is possible
to saturate the model-independent values for $S_f$ and $a_{\rm SL}$ 
shown in fig.~\ref{fig:model_independent}.

In the left and middle plots of fig.~\ref{fig:SUSY}, we show the correlation 
between $d_n$ and $S_f$ as well as $d_n$ and $a_{\rm SL}$.

As discussed in the previous sections, the CPV observables $S_f$ and $a_{\rm SL}$
are generated by the imaginary part of the $D^0-\bar D^0$ mixing amplitude
${\rm Im} M_{12}\sim{\rm Im}\left[(\delta^{L}_{u})_{21}(\delta^{R}_{u})_{21}\right]$.
At the same time, also the hadronic EDMs are generated by means of the up-quark (C)EDM
$d^{(c)}_{u}\sim{\rm Im}M^{\star}_{12}\sim{\rm Im}\left[(\delta^{L}_{u})_{21}(\delta^{R}_{u})^{*}_{21}\right]$.

Examples of relevant Feynman diagrams contributing to $D^0 - \bar D^0$ mixing and to the
up quark (C)EDM in SUSY alignment models are shown in fig.~\ref{fig:diagrams}.

Even if the CP violating source is the same, $d_n$ and $S_f$ cannot be exactly correlated.
The reason is twofold: i) $d_n\sim{\rm Im}M^{\star}_{12}$ while the relevant phase for $S_f$ is $\phi=\textnormal{Arg}(q/p)$ with $q/p$ defined in eq.~(\ref{q/p}), ii) while $d^{(c)}_{u}$ is sensitive to $A_c$ (see eq.~(\ref{Eq:edm_u_gluino})),
${\rm Im}M^{\star}_{12}$ is not (see eq.~(\ref{M^D_12})). Even if the natural value for $A_c$
is $A_c \sim m_{\tilde g},\tilde m_Q$, there are corners of the SUSY parameter space
where $A_c \ll m_{\tilde g},\tilde m_Q$ (remember that within the CMSSM-like spectrum, which we assume, $A_c\approx 0.65~A_0-2.8~M_{1/2}$).

However, interestingly enough, large values for $S_f$ and $a_{\rm SL}$
necessarily imply a lower bound for the neutron EDM $d_n\gtrsim 10^{-(28-29)}e$~cm, that
is an experimentally interesting level for the expected future experimental resolutions.
\footnote{Similarly, in ref.~\cite{Altmannshofer:2008hc} we pointed out that, in the context
of the flavour blind MSSM (FBMSSM), large (non-standard) CP violating effects for $b \to s$
transitions, like the CP asymmetries in $B_d \to \phi K_S$ and $b \to s \gamma$, unambiguously predict a lower bound for the electron and neutron EDMs (as well as for the mercury EDM) in
the reach of the future experimental sensitivities.}

Similarly, according to eq.~(\ref{Eq:dHg}), it turns out that the corresponding lower
bound for $d_{\rm Hg}$ is $d_{\rm Hg}\gtrsim 10^{-(30-31)}e$~cm.

In summary the most peculiar predictions for the SUSY Abelian models of table~\ref{QSA} are:
\begin{itemize}
\item Natural solution of the SUSY flavour problem thanks to small (most probably undetectable)
      effects in the down quark sector, i.e. in $K^0-\bar K^0$, $B^0-\bar B^0$ and
      $B^0_s-\bar B^0_s$ mixings.\footnote{An exception to these findings arises, for example,
      in the context of the Abelian flavour model proposed in ref.~\cite{Agashe:2003rj}, which
      does not belong to the class of the Abelian flavour models of table~\ref{QSA}. As thoroughly
      discussed in ref.~\cite{Altmannshofer:2009ne}, the model of ref.~\cite{Agashe:2003rj}
      predicts, in addition to large NP effects in $D^0-\bar D^0$ mixing, also large NP effects
      for $b \to s$ transitions.}
\item Experimentally visible CP violating effects in $D^0-\bar D^0$ mixing, as the time
      dependent CP asymmetry in decays to CP eigenstates $S_f$ and the
      semileptonic asymmetry $a_{\rm SL}$.
\item Large values for the hadronic EDMs, in the reach of the future experimental sensitivities,
      generated by the up-quark (C)EDM. Hence, a correlated study of several hadronic EDMs, with
      different sensitivity to the up-quark (C)EDM would provide a crucial tool to probe SUSY
      alignment models.
\item A lower bound for the EDM of hadronic systems (like the neutron EDM and the mercury EDM),
      in the reach of future experimental sensitivities, for given large (non-standard) values
      of $S_f$ and $a_{\rm SL}$.
\end{itemize}

\section{Conclusions}

Within the SM, CP violating effects in $D$ meson systems are predicted to be highly
suppressed at the level of ${\cal O}((V_{cb}V_{ub})/(V_{cs}V_{us}))\sim10^{-3}$.
Therefore, any experimental evidence for CP violation in $D^0-\bar D^0$ mixing
above the per mill level would unambiguously point towards a NP effect.

On the other hand, within supersymmetric scenarios like the popular alignment models,
large non-standard effects in $D^0-\bar D^0$ mixing naturally arise. At the same time,
the NP content of FCNC and/or CP violating observables related to the $K$, $B_d$ and
often also $B_s$ systems is small enough and thus hardly detectable because of
irreducible hadronic uncertainties.

In this letter we have demonstrated that, within alignment models, large CP violating
effects in $D^0-\bar D^0$ mixing would unambiguously predict a lower bound for the EDMs
of hadronic systems, like the neutron EDM and the mercury EDM, well within the future
experimental sensitivities.

As a byproduct of our analysis, we also studied the correlation between two CP violating
observables in $D$ mixing: the time dependent CP asymmetry in decays to CP eigenstates
$S_f$ and the semileptonic asymmetry $a_{\rm SL}$ both model independently 
and, for the first time, within supersymmetry.

The above physics case provides a splendid example of the complementarity existing
among different low energy observables in shedding light on the NP theory at work,
if any. The simultaneous evidence of non-standard CP violating effects in the $D$
meson system together with the evidence of non vanishing hadronic EDMs would strongly
support the idea of SUSY alignment models and disfavor gauge-mediated SUSY breaking
models, SUSY models with MFV and non-Abelian SUSY flavour models.

Viceversa, if nature has chosen supersymmetry to be the completion of the SM theory
operating at the TeV scale and if alignment models are responsible for the solution
of the SUSY flavour problem, it is uniquely predicted that sizable NP effects
might appear in CP violating modes of the $D$ meson system (but most probably not
in the $B_s$ system) and in the hadronic EDMs.

Hence, any experimental effort aiming to improve their current resolutions are highly
desired and hardly overemphasized.

\textit{Acknowledgments:}
We thank Ikaros Bigi for illuminating discussions on the phase $\phi$.
This work has been supported in part by the Cluster of Excellence ``Origin and Structure
of the Universe'' and by the German Bundesministerium f{\"u}r Bildung und Forschung under
contract 05H09WOE.


\end{document}